\documentclass[prl,aps,twocolumn,showpacs,amsmath,amssymb]{revtex4}
\usepackage{epsf}

\begin{document}

\title{Spin dynamics in the regime of hopping conductivity}
\author{I.\,S.~Lyubinskiy, A.\,P.~Dmitriev, and V.\,Yu.~Kachorovskii}

\affiliation{A.F.~Ioffe Physical-Technical Institute, 26 Polytechnicheskaya street, Saint Petersburg, 194021, Russia}

\date{\today}

\pacs{
71.70.Ej, %spin-orbit coupling in condensed matter
73.63.-b, %low-dimensional structures, electrical properties
75.40.Gb, %spin diffusion
85.75.-d  %spintronics
}

\begin{abstract}
We consider spin dynamics in the impurity band of a
semiconductor with spin-split spectrum. Due to the splitting,  phonon-assisted hops from one impurity to another are
accompanied by  rotation of the electron spin, which leads to spin relaxation. The system is strongly inhomogeneous because of exponential variation of hopping times. However, at very small couplings an   electron   diffuses over  a distance exceeding the characteristic scale of the inhomogeneity during the time of  spin relaxation, so one can introduce  an averaged spin relaxation rate.  At larger values of coupling 
%the relaxation  can not be described by averaged rate. In this %case 
the system is effectively divided into two subsystems: the one where  relaxation is very fast and another one where relaxation is rather slow. In this case, spin decays due to escape of the electrons from one subsystem to another.
%regions where relaxation is slow to the regions where relaxation %is very fast. The spin dynamics is controlled rather by the %escape time than by the relaxation time on the later regions. 
As a result, the spin dynamics is non-exponential  and hardly depends on spin-orbit coupling. 
\end{abstract}

\maketitle

Spin dynamics in  semiconductors
  has attracted much attention in the last decades (for review, see Ref.~\cite{aw}).
  A great number of publications were devoted to
 the study of  the spin relaxation  in  the metallic  regime, where it is usually dominated by   the 
Dyakonov-Perel mechanism. According to this mechanism, the spin relaxation rate is proportional to the diffusion coefficient \cite{dp}. Recently, the idea was put forward \cite{sh} that this proportionality goes  beyond the metallic regime and  stays valid  in the regime of  hopping conductivity:  
 \begin{equation}
 {1}/{\tau^{(1)}_S} \sim {D}/{L_S^2}.
 \label{MS}
 \end{equation}
  Here $\tau^{(1) }_S$ is the spin relaxation time, 
$L_S \sim v/\Omega,$  $v$ is the tunnelling  velocity,
 and  $\Omega$   is the typical value of spin precession frequency for an electron moving under the barrier.

In this paper we discuss spin dynamics in the hopping regime and show that a number of  spin relaxation mechanisms are realized in addition to the one discussed in Ref.~\cite{sh}. 
The hopping regime takes place at small temperatures $T \ll W$ (here $W$ is the width of the impurity band). In the opposite limit $T \gg W$ the transport is governed by activation to the mobility edge.  
However, methodologically it is convenient  to  start from the modelling case when  $T \gg W$ but transitions to the conduction band are forbidden.
This model is rather simple because the probability of a hop does not depend on temperature. On the other hand, it captures basic physics of the problem and  
can be easily generalized to the real situation. 
We   neglect
 electron-electron interaction and assume that the distance between impurities   is much larger than the Bohr radius $a$:   $na^d \ll 1$ (here $n$ is the impurity concentration). We also assume that $L_S \gg a$ which is usually the case.
% in real semiconductor systems. 

Due to the spin-orbit coupling,  phonon-assisted hops 
are accompanied by  rotation
of the electron spin.   The rotation angle can be written as 
$\boldsymbol {\varphi} = \boldsymbol {\varphi}_0+ \delta\boldsymbol {\varphi},
 $ where  $\boldsymbol {\varphi}_0$ corresponds to the quasiclassical approximation, which implies that electron moves under barrier 
along the straight line connecting the impurities and  $\delta\boldsymbol {\varphi}$ is the  correction which is due to quantum uncertainty  of the trajectory (this correction cannot be neglected when we consider non-typical hops over the distances of the order of several $a$). 
The angle $\boldsymbol {\varphi}_0$ is given by
$\boldsymbol{\varphi_0} =\boldsymbol{\Omega}(\mathbf v) t,$
where
$\boldsymbol{\Omega}(\mathbf v)$ is proportional to spin-orbit spectrum
splitting, $\mathbf v =  \mathbf n \hbar/ma$ is the
tunnelling velocity, $t=  r/v$, $a$ is the Bohr radius, $m$
is the electron mass, $r$ is the hopping length, and  $\mathbf n $ is the unit vector in the direction of the hop. 
In the 3D case $\Omega_x \sim v_x(v_y^2-v_z^2)$
and the other components are given by cyclical transmutations \cite{dress}. In the 2D case $\boldsymbol{\Omega} = \mathbf a v_x +\mathbf b v_y$, where constant vectors $ \mathbf a $ and  $\mathbf b $  depend on the quantum well orientation and the degree of asymmetry of the confining potential \cite{rashba,dpqwell}.
The characteristic value of $\boldsymbol{\varphi_0}$ is given by
\begin{equation}
\varphi_0 \sim  r/L_S.
\label{varphi}
\end{equation}

Random hops between impurities lead to spin relaxation.
Let us  assume that at  $t=0$ the homogeneous distribution of the spin density $\mathbf S_0$ is created in the system (for example, by means of optical orientation) and study the dynamics of the averaged spin density $\mathbf S(t)$. 
The system under discussion is strongly inhomogeneous due to exponential variation of hopping times. 
Indeed, the waiting time $\tau_{ij}$  of a hop between impurities $i$ and $j$ varies exponentially with the distance $L_{ij}$ between impurities: $\tau_{ij} = \tau_0 \exp({2L_{ij}/a}),$ where $\tau_0$ is the waiting time for the impurities separated by the distance of the order of $a$ (here we took into account that $T \gg W$).
%In this letter we will show that, due to inhomogeneity, the system effectively divides into clusters with exponentially varying local spin relaxation rates. At sufficiently small spin-orbit coupling, when the relaxation is slow, an electron has enough time to visit many such clusters. In this case one can introduce an averaged spin relaxation rate $1/\tau_S$ so that $\mathbf S(t)=\mathbf S_0\exp(-t/\tau_S)$.  In the opposite limit spin relaxation is controlled by escapes of electrons from the clusters where relaxation is slow to those where relaxation is fast. One of our main results is that in the later regime the   spin dynamics hardly depends on the spin orbit coupling strength.
%Now, let us describe the procedure of dividing the system into clusters.  Following Ref.~\cite{ef}, 
We assume \cite{ef}
that  impurities $i$ and $j$ are effectively connected if $\tau_{ij}<t,$ or, equivalently, $$L_{ij} <L(t)=a/2\ln(t/\tau_0),$$ and disconnected if $\tau_{ij}>t$ ($L_{ij}>L(t)$).  For $L(t)$ larger than  a certain critical length $L_h \sim n^{-1/d}$, the connected
impurities  form infinite cluster. The fraction of impurities $P(\xi)$ belonging to the infinite cluster increases with increasing of $t$ as shown in Fig.1.   Here
\begin{equation}
\xi = \xi(t) = L(t)/L_h = (a/2L_h)  \ln (t/\tau_0).
\end{equation}
Near the critical point ($L(t)=L_h$) the share of  impurities belonging to the infinite cluster $P$ behaves as a power function: $P = (\xi - 1)^{\beta},$ (where $\beta$ is critical index), and with the further increase of  $\xi$ it  is quickly saturated \cite{ef}. 

The exponential regime of the spin relaxation $\mathbf S(t)=\mathbf S_0\exp(-t/\tau_S)$ with an averaged rate $1/\tau_S$  is realized at sufficiently small spin-orbit coupling, 
when an electron has enough time to visit almost any possible impurity configuration during the time $\tau_S$.   
%relaxation is slow, such a saturation takes place  before %electron spin relaxes. %an electron has enough time to visit %many such clusters. 
%In this case one can introduce an averaged spin relaxation rate %$1/\tau_S$ so that $\mathbf S(t)=\mathbf S_0\exp(-t/\tau_S)$.
%This regime is realized when
%At small spin-orbit coupling  
This condition is satisfied when
\begin{equation}
1-P[\xi(\tau_S)] \ll 1.
\label{us}
\end{equation}
%This condition is necessary for self-averaging  of the spin relaxation rate. 
There are two contributions to $1/\tau_S$. The first one is due to electron diffusion on the scales larger than the scale of inhomogeneity. This contribution is given by Eq.~\eqref{MS} \cite{sh}.
Another contribution, neglected in Ref.~\cite{sh}, comes from the spin
relaxation on non-typical clusters with the distances between  impurities on the order of several $a$. The most likely non-typical configurations of impurities are pairs and triangles. %
%An electron captured on such a complex makes many hops before it %leaves it. The shift in space due to these hops is limited by %the size of the complex, while the decrease of its spin is %proportional to the number of hops. Therefore, the electron does %not participate in diffusion, but these complexes might give %essential contribution to the spin relaxation.
%
An electron captured on  a non-typical complex   makes many hops in a small region of space before it leaves it. The shift in space due to these hops is limited by the size of the complex, while the deviation  of its spin increases with the number of hops.
%These hops do not lead to a significant shift in the space and,  %
Therefore,  such complexes    do not  contribute to the diffusion process but they
%. However, due to frequent hops, the non-typical complexes 
might give essential contribution to the spin relaxation.
%
%
%
%
%
%The probability to find such configurations is low, so they do %not give essential contribution to the electron diffusion. 
%However, electrons captured on such complexes makes frequent hops and, as a consequence, their spin might quickly relax.   
%
 %
%However, due to frequent hops, such impurity configurations might give essential contribution to the spin relaxation, even though 
%However, due to frequent hops, such impurity configurations might give essential contribution to the spin relaxation, even though the probability to find such configurations is low.  

The most probable non-typical configuration is a pair of impurities. 
In quasi-classical approximation the contribution of pairs to spin relaxation is equal to zero, because the rotation of the spin during the hop from one impurity to another is compensated by the rotation during the hope in the opposite direction. However, when the distance between impurities is of the order of $a$, the deviation of the trajectory from the quasi-classical one due to quantum uncertainty is of the order of the size of the quasiclassical trajectory itself. 
This implies that the correction $\delta{\varphi}$ to the rotation angle is comparable with 
${\varphi_0}$ 
(i.e. $\delta{\varphi} \sim \varphi_0 \sim a/L_S\ll 1 $). This is indeed the case in 3D semiconductors.
In the 2D case the situation is more subtle. In this case the rotation angle in the first order of $1/L_S$ depends only on the shift of the electron in space $\boldsymbol{\varphi} = \int \boldsymbol{\Omega} dt = \mathbf a r_x + \mathbf b r_y$ (here the integration is taken over the electron trajectory and we took into account that small rotations commute) and does not depend on a particular trajectory. As a result, the correction to the main contribution appears only in the next orders in $1/L_S$: $\delta{\varphi} \sim (a/L_S)^m$, where $m \ge 2$. One can show that  $m=2$ for the components of the spin which are perpendicular to $[\mathbf a \times \mathbf b]$ and $m=3$ for the parallel  component.
The exact value of this correction depends on the wave vector of the phonon that was emitted or absorbed during the hop \cite{fnote1}. Therefore, the rotation angle is a random quantity. 

  In general case,  the angle of the rotation after passing the closed trajectory (that includes the hop from the first impurity to the second one and the hop in the opposite direction) can be written as:
\begin{equation}
\delta{\varphi} \sim (a/L_S)^m,
\end{equation}
where $m=1$ in 3D case and $m=2,3$ in 2D case.

After   rotation by the angle $\delta{\varphi}$,  spin projection on the original direction decreases  on the average by $\delta{\varphi}^2$. So, one can expect that a   relaxation rate on a single pair is given by $ (a/L_S)^{2m} (1/\tau_0)$. However, this estimate does not take into account  the effects of level repulsion. The repulsion energy is on the order of $\Delta E \sim (e^2/\kappa a) e^{-r/a}$.  A non-typical pair can participate in the spin relaxation only if $\Delta E \le T$. Therefore, the minimal distance between impurities in a non-typical pair is given by  $r^* \ge a \ln(\kappa a /T e^2)$. The hop waiting time for such a pair is given by $\tau_0\exp(2r^*/a) \sim \tau_0 (e^2/T \kappa a)^2$, this leads to an additional factor $(T \kappa a/e^2)^2 \ll 1$ in the spin relaxation rate. 
As a result, the rate of the spin relaxation on a single non-typical pair reads
\begin{equation}
1/\tau_S^* \sim (a/L_S)^{2m} (1/\tau_0)(T \kappa a/e^2)^2.
\end{equation}
The contribution of non-typical pairs to the spin relaxation rate is proportional to $1/\tau_S^*$
%
%the spin relaxation rate on a single pair 
and to the share of the time that electron spends on non-typical pairs. This share is proportional to the ratio of the  concentration of non-typical pairs $n^*$ to the total impurity concentration: $n^*/n \sim (a/L_h)^d$. 
Hence, the corresponding relaxation rate is given by
\begin{equation}
{1}/{\tau^{(2)}_S} \sim  (a/L_h)^d
\left({a}/{L_S}\right)^{2m}({1}/{\tau_0})(T \kappa a/e^2)^2. \label{ave}
\end{equation}

Both mechanisms described above are additive and the spin relaxation rate is:
\begin{equation}\label{tauS}
  1/\tau_S =1/\tau^{(1)}_S+1/\tau^{(2)}_S.
\end{equation}
Let us compare these two mechanisms.
In the case under discussion, when $T\gg W$, the diffusion coefficient entering  Eq.~\eqref{MS} is determined by the hops over the critical  length: $D\sim\exp(-2L_h/a)$ \cite{ef}. Therefore, $1/\tau_S^{(1)}$  decreases exponentially with the average distance between impurities, while  $1/\tau_S^{(2)}$ decreases according to slow power law  $1/L_h^d$ (see Eq.~\eqref{ave}).   In the 3D case the dependence on $L_S$ in  Eq.~\eqref{ave} is the same as in  Eq.~\eqref{MS}. Therefore, in the 3D case, $1/\tau_{S}^{(2)} \gg 1/\tau_{S}^{(1)}$ (since we assumed that  $L_h \gg a$).
In the 2D case, $1/\tau^{(2)}_S$ decreases with $L_S$  faster than $1/\tau_{S}^{(1)}$. Thus, at very small  spin-orbit coupling the main contribution is given by Eq.~\eqref{MS}. However, 
%With the increasing of the  coupling the transition to the %regime described by Eq.~\eqref{ave} takes place. In the limit %$L_h \gg a$, 
the crossover to the regime of non-typical pairs  takes place already at exponentially small value of the coupling.

In deriving   Eq.~\eqref{ave} we assumed that during the process of spin relaxation an electron has enough time to  visit many non-typical pairs which are separated by a    distance  on the order of $[n^{*}]^{-1/d}=L_h(L_h/a)$  (this is the length that plays the role of inhomogeneity scale in this regime). 
With the increasing of the strength of spin-orbit coupling this condition fails because spin relaxation time decreases and 
there appear finite  clusters 
that are effectively separated from the rest of the system on the time scale on the order of $\tau_S$. Since the concentration of non-typical pairs $n^{*}$  is small, most of the isolated clusters do not contain them. 
 % The  mechanism proposed in Ref.~\cite{sh} also does not work %for these clusters, because the diffusion coefficient
%for the finite area is  equal to zero.  
This means that the spin of  electrons captured  on the isolated
clusters does not relax until they escape
 to the infinite cluster where the relaxation is much faster (it  is still determined by  Eq.~\eqref{tauS}).  Therefore, the infinite cluster can be considered as a "black hole" for electron spin. The  spin dynamics  in this regime is  given by
\begin{equation}
\mathbf S(t)=\mathbf S_0 [1-P\left(t\right)],
\label{cl}
\end{equation}
where $ P(t)=P[{L(t)}/{L_h}]= P[( {a}/{2L_h})\ln({t}/{\tau_0})]$. Hence, the spin polarization is a function of a  single
dimensionless parameter $(a/2L_h) \ln(t/\tau_0).$
 It worth noting that the dynamics of the spin relaxation in this
 regime does not depend on the  strength of spin-orbit coupling.
\begin{figure}[ht!]
\vspace{25mm} 
\hspace{-20mm}
\leavevmode \epsfxsize=5cm
\centering{\epsfbox{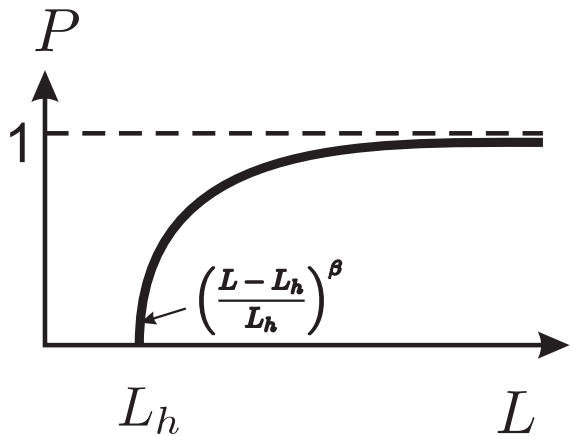}}
\caption{Fraction of impurities belonging to the infinite cluster
as a function of the maximal distance $L$ between connected
impurities.  In the vicinity of the critical length $L_h,$
function $P[L/L_h]$ increases as $[L/L_h-1]^{\beta},$ where
$\beta$ is a critical index.}
\end{figure}
%
%Now we can formulate a quantitative criteria of the %applicability of the regime of optimal pairs.

With the further increase of the  coupling, the spin
relaxation on the finite clusters comes into play. Finally, at
extremely high  coupling the spin rotates by a large angle  after
one typical hop. In this case, the spin relaxes on any finite
clusters except isolated impurities 
%(small
%clusters containing one or two impurities) 
separated from the rest
of the system by  distances larger than  $L(t).$ 
The spin at the  moment $t$ is proportional to the number of
electrons captured on such "traps":
\begin{equation}
\mathbf S(t) = \mathbf S_0 \exp \left[-  C_d\left(\frac{a}{L_h} \ln
\frac{t}{\tau_0}\right)^d \right],
\label{isol}
\end{equation}
where $d=2,3$ is dimensionality of the system and $C_d \sim 1$ is a numerical coefficient. It is seen  from
this equation that just as in the case of intermediate coupling
 the spin dynamics
does not depend on the spin-orbit strength and determined by the
only parameter $(a/L_h) \ln(t/\tau_0).$ 

Next, we discuss qualitatively the generalization of the approach discussed above to the case of small temperatures. In this case the waiting time is given by $\tau_{ij} = \tau_0 \exp(L_{ij}/a+ E_{ij}/T)$, where $E_{ij}$ is the energy difference between states $i$ and $j$. The minimization of this exponent yields the optimal values of hopping length $L_{ij}\sim L_h(T)$ and hopping energy $E_{ij}\sim E_h(T)$ which are given by \cite{ef}:
\begin{equation}
\frac{E_h(T)}{T} \sim \frac{L_h(T)}{a} \sim (g T a^d)^{-1/(d+1)}
\end{equation}
(these equations are applicable when $T < W a/L_h$).
The optimal length $L_h(T)$ is of the order of the average distance between impurities lying in the energy band of width $E_h(T)$ (optimal energy band). The states outside the band of width $E_h(T)$ near the Fermi level  are not accessible to polarized electrons.  As a consequence, the equations derived   above are still valid with two modifications. First, one should make a replacement $L_h \to L_h(T)$. Second, the effective dimension of the system should be increased $d \to d+1$, which takes into account the motion of the electron along energy axis.

To conclude, the theory of the spin relaxation of the electrons in the impurity band was developed and a number of different relaxation regimes were predicted. At weak spin-orbit couplings
spin relaxation is exponential with the rate of spin relaxation given by two contributions:
the first one comes from the relaxation on nontypical pairs of impurities and the second one is due to electron diffusion on the scales larger than the scale of inhomogeneity of the system. At stronger couplings, spin relaxation is due to escapes of the electrons from finite clusters to the infinite one. In this case, the law of spin relaxation does not depend on the spin-orbit coupling. At very large couplings the spin relaxation is due to escape from the spin traps.

We are grateful to M.I.~Dyakonov, V.I.~Perel, and B.I.~Shklovskii  for useful discussions.

This work has been supported by  RFBR, by  grants of the
RAS, by a grant of the Russian Scientific School 2192.2003.2, and a
grant of the foundation "Dinasty"-ICFPM.


\begin{thebibliography}{99}

\bibitem{aw}
{\it Semiconductor Spintronics and Quantum Computation}, edited by
D.\,D. Awschalom, D. Loss, and N. Samarth (Springer-Verlag, Berlin,
2002).

\bibitem{dp}
M. I. Dyakonov and V. I. Perel', Fiz. Tverd. Tela  {\bf 13}, 3581 (1971) [Sov. Phys. Solid State {\bf 13}, 3023 (1972)].
\bibitem{sh} B.I. Shklovskii, Phys. Rev. B {\bf 73}, 193201 (2006).

\bibitem{dress}
G. Dresselhaus, Phys. Rev. {\bf 100}, 580 (1955).

\bibitem{rashba}
Yu. A. Bychkov and E. I. Rashba, Pis'ma Zh. Eksp. Teor. Fiz. {\bf 39}, 66 (1984) [JETP Lett. {\bf 39}, 78 (1984)].

\bibitem{dpqwell}
M. I. Dyakonov and V. Yu. Kachorovskii, Fiz. Tekh. Poluprovodn.  {\bf 20}, 178 (1986) [Sov. Phys. Semicond. {\bf 20}, 110 (1986)].


\bibitem {ef} B. I. Shklovskii and A. L. Efros, {\it Electronic Properties of Doped Semoconductors} (Springer-Verlag, Berlin, 1984).

\bibitem {fnote1}
The rigorous approach to the derivation of $\delta{\varphi}$ implies calculation of the matrix element that describes phonon-assisted hop from one impurity to another. Such calculation shows that in the limit $qa \le 1$ (where $\mathbf q$ is the phonon wave-vector), that is most relevant in the realistic  semiconductor structures, the characteristic value of $\delta{\varphi}$ does not depend on the parameter $qa$, but depends on the direction of the vector $\mathbf q$.


\end{thebibliography}
\end{document}